\documentclass[preprintnumbers,amsmath,amssymbm,prd]{revtex4}
\usepackage{epsfig}
\usepackage{graphicx}

\begin{document}
\title{Slowly decaying resonances of charged massive scalar fields in the Reissner-Nordstr\"om black-hole spacetime}
\author{Shahar Hod}
\affiliation{The Ruppin Academic Center, Emeq Hefer 40250, Israel}
\affiliation{ }
\affiliation{The Hadassah Institute, Jerusalem 91010, Israel}
\date{\today}

\begin{abstract}
\ \ \ We determine the characteristic timescales associated with the
linearized relaxation dynamics of the composed
Reissner-Nordstr\"om-black-hole-charged-massive-scalar-field system.
To that end, the quasinormal resonant frequencies
$\{\omega_n(\mu,q,M,Q)\}_{n=0}^{n=\infty}$ which characterize the
dynamics of a charged scalar field of mass $\mu$ and charge coupling
constant $q$ in the charged Reissner-Nordstr\"om black-hole
spacetime of mass $M$ and electric charge $Q$ are determined {\it
analytically} in the eikonal regime $1\ll M\mu<qQ$. Interestingly,
we find that, for a given value of the dimensionless black-hole
electric charge $Q/M$, the imaginary part of the resonant
oscillation frequency is a monotonically {\it decreasing} function
of the dimensionless ratio $\mu/q$. In particular, it is shown that
the quasinormal resonance spectrum is characterized by the
asymptotic behavior $\Im\omega\to0$ in the limiting case $M\mu\to
qQ$. This intriguing finding implies that the composed
Reissner-Nordstr\"om-black-hole-charged-massive-scalar-field system
is characterized by extremely long relaxation times
$\tau_{\text{relax}}\equiv 1/\Im\omega\to\infty$ in the $M\mu/qQ\to
1^-$ limit.
\end{abstract}
\bigskip
\maketitle

\section{Introduction}

Wheeler's `no-hair' conjecture \cite{Whee,Car} presents a simple
physical picture according to which asymptotically flat black holes
in Einstein's theory of gravitation cannot support static matter
fields outside their horizons. Interestingly, various no-hair
theorems \cite{Whee,Car,Bek1,Chas,BekVec,Hart,Nun,Hod11} provide
strong support for the general validity of Wheeler's famous
conjecture. It is therefore expected that fundamental matter fields
that propagate in a static black-hole spacetime would eventually be
absorbed by the black hole or be scattered away to infinity. For
rotating black-hole spacetimes, a third possibility also exists
\cite{Hodrc,HerR}: thanks to the intriguing physical mechanism of
superradiant scattering of bosonic fields in spinning black-hole
spacetimes, these black holes can support stationary (rather than
static) linearized bound-state massive scalar configurations in
their external regions \cite{Hodrc}. Moreover, as explicitly shown
in \cite{HerR}, rotating black holes can also support genuine
bosonic hair (that is, non-linear stationary bosonic field
configurations) in their external regions. These rotating hairy
black-hole-bosonic-field configurations \cite{Hodrc,HerR} provide
explicit counterexamples to the no-hair conjecture in asymptotically
flat non-static spacetimes.

It is important to stress the fact that the elegant no-hair theorems
\cite{Whee,Car,Bek1,Chas,BekVec,Hart,Nun,Hod11}, which are valid for
asymptotically flat static black-hole configurations, say nothing
about the {\it timescale} associated with the dynamical process of
black-hole hair shedding \cite{Notetim}. This characteristic
relaxation timescale, $\tau_{\text{relax}}$, will be the main focus
of the present study.

The dynamics of fundamental matter and radiation fields in
black-hole spacetimes are characterized by {\it damped} quasinormal
oscillations of the form $e^{-i\omega t}$
\cite{QNMs,Notetails,Tails}. These exponentially decaying
oscillations are characterized by complex quasinormal resonant
frequencies $\{\omega_n\}_{n=0}^{n=\infty}$ whose values depend on
the physical parameters (such as mass, charge, angular momentum, and
intrinsic spin) of the composed black-hole-field system. In accord
with Wheeler's no-hair conjecture \cite{Whee,Car}, these
characteristic damped oscillations reflect the gradual decay of the
fields in the external regions of the black-hole spacetimes. In
particular, the characteristic timescale associated with the
relaxation dynamics of an external field in a black-hole spacetime
is determined by the imaginary part of the fundamental (least
damped) quasinormal resonant frequency which characterizes the
composed black-hole-field system:
\begin{equation}\label{Eq1}
\tau_{\text{relax}}\equiv 1/\Im\omega_0\  .
\end{equation}

The main goal of the present paper is to determine the
characteristic relaxation timescales, $\tau_{\text{relax}}$,
associated with the relaxation dynamics of charged massive scalar
fields in the charged Reissner-Nordstr\"om (RN) black-hole
spacetime. To that end, we shall explore below the quasinormal
resonance spectrum which characterizes the linearized relaxation
dynamics \cite{Notenhc,Hodth} of the composed
RN-black-hole-charged-massive-scalar-field system. As we shall show
below, the characteristic quasinormal resonances of this composed
black-hole-field system can be studied {\it analytically} in the
eikonal regime \cite{Noteunit}
\begin{equation}\label{Eq2}
1\ll M\mu<qQ\  ,
\end{equation}
where $\{q,\mu\}$ are respectively the charge coupling constant and
proper mass of the field, and $\{M,Q\}$ are respectively the mass
and electric charge of the RN black hole. In particular, below we
shall reveal the interesting fact that this composed
RN-black-hole-charged-massive-scalar-field system is characterized
by extremely long dynamical relaxation times,
$\tau_{\text{relax}}\equiv 1/\Im\omega_0\to\infty$, in the limiting
case $M\mu/qQ\to 1^-$ \cite{Notekj,Pjk}.

\section{Description of the system}

We shall analyze the quasinormal resonance spectrum which
characterizes the linearized relaxation dynamics of a scalar field
$\Psi$ of mass $\mu$ and charge coupling constant $q$ \cite{Noteqm}
in the spacetime of a Reissner-Nordstr\"om black hole of mass $M$
and electric charged $Q$ \cite{Notegen}. The curved black-hole
spacetime is described by the line element \cite{Chan}
\begin{equation}\label{Eq3}
ds^2=-f(r)dt^2+{1\over{f(r)}}dr^2+r^2(d\theta^2+\sin^2\theta
d\phi^2)\ ,
\end{equation}
where
\begin{equation}\label{Eq4}
f(r)=1-{{2M}\over{r}}+{{Q^2}\over{r^2}}\  .
\end{equation}
The zeros of the radial function $f(r)$,
\begin{equation}\label{Eq5}
r_{\pm}=M\pm(M^2-Q^2)^{1/2}\  ,
\end{equation}
determine the horizon radii of the charged RN black hole.

The familiar Klein-Gordon wave equation
\cite{HodPirpam,Stro,HodCQG2,Hodch1,Hodch2,Konf,Ric}
\begin{equation}\label{Eq6}
[(\nabla^\nu-iqA^\nu)(\nabla_{\nu}-iqA_{\nu})-\mu^2]\Psi=0\
\end{equation}
determines the linearized dynamics of the charged massive scalar
field in the curved RN black-hole spacetime, where
$A_{\nu}=-\delta_{\nu}^{0}{Q/r}$ is the electromagnetic potential of
the charged black hole. Substituting the field decomposition
\cite{Noteom}
\begin{equation}\label{Eq7}
\Psi(t,r,\theta,\phi)=\int\sum_{lm}e^{im\phi}S_{lm}(\theta)R_{lm}(r;\omega)e^{-i\omega
t} d\omega\ ,
\end{equation}
into the Klein-Gordon wave equation (\ref{Eq6}), and using the
black-hole metric function (\ref{Eq4}), one obtains
\cite{HodPirpam,Stro,HodCQG2} two ordinary differential equations of
the confluent Heun type \cite{Heun,Abram} for the eigenfunctions
$R(r)$ and $S(\theta)$ which respectively describe the radial and
angular behaviors of the charged massive scalar field in the curved
black-hole spacetime.

The ordinary differential equation which determines the spatial
behavior of the radial eigenfunction $R(r)$ is given by
\cite{HodPirpam,Stro,HodCQG2}
\begin{equation}\label{Eq8}
\Delta{{d} \over{dr}}\Big(\Delta{{dR}\over{dr}}\Big)+UR=0\ ,
\end{equation}
where $\Delta=r^2f(r)$, and
\begin{equation}\label{Eq9}
U=(\omega r^2-qQr)^2 -\Delta(\mu^2r^2+K_l)\  .
\end{equation}
Here $K_l=l(l+1)$ (with $l\geq |m|$) is the characteristic
eigenvalue of the angular eigenfunction $S_{lm}(\theta)$
\cite{Heun,Abram,Notesy}.

Defining the ``tortoise" radial coordinate $y$ by the differential
relation
\begin{equation}\label{Eq10}
dy={{dr}\over{f(r)}}\  ,
\end{equation}
and using the new radial eigenfunction
\begin{equation}\label{Eq11}
\psi=rR\  ,
\end{equation}
one can transform the radial equation (\ref{Eq8}) into the more
familiar form
\begin{equation}\label{Eq12}
{{d^2\psi}\over{dy^2}}+V\psi=0\
\end{equation}
of a Schr\"odinger-like ordinary differential equation, where the
effective radial potential in (\ref{Eq12}) is given by
\begin{equation}\label{Eq13}
V=V(r;M,Q,\omega,q,\mu,l)=\Big(\omega-{{qQ}\over{r}}\Big)^2-f(r)H(r)\
\end{equation}
with
\begin{equation}\label{Eq14}
H(r;M,Q,\mu,l)=
\mu^2+{{l(l+1)}\over{r^2}}+{{2M}\over{r^3}}-{{2Q^2}\over{r^4}}\ .
\end{equation}

The quasinormal resonant frequencies
$\{\omega_n(M,Q,\mu,q,l)\}_{n=0}^{n=\infty}$, which characterize the
linearized relaxation dynamics of the charged scalar field in the
charged black-hole spacetime, are determined by imposing on the
Schr\"odinger-like wave equation (\ref{Eq12}) the physically
motivated boundary conditions of purely ingoing waves at the
black-hole horizon and purely outgoing waves at spatial infinity
\cite{Detw}. That is,
\begin{equation}\label{Eq15}
\psi \sim
\begin{cases}
e^{-i (\omega-qQ/r_+)y} & \text{ as\ \ \ } r\rightarrow r_+\ \
(y\rightarrow -\infty)\ ; \\ y^{-iqQ}e^{i\sqrt{\omega^2-\mu^2} y} &
\text{ as\ \ \ } r\rightarrow\infty\ \ (y\rightarrow \infty)\  .
\end{cases}
\end{equation}

In the next section we shall study {\it analytically} the
quasinormal resonance spectrum
$\{\omega_n(M,Q,\mu,q,l)\}_{n=0}^{n=\infty}$ which characterizes the
relaxation dynamics of the composed
Reissner-Nordstr\"om-black-hole-charged-massive-scalar-field system
in the eikonal large-mass regime (\ref{Eq2}).

\section{The quasinormal resonance spectrum of the composed
Reissner-Nordstr\"om-black-hole-charged-massive-scalar-field system}

In the present section we shall perform a WKB analysis in order to
determine the complex resonant frequencies which characterize the
composed
Reissner-Nordstr\"om-black-hole-charged-massive-scalar-field system
in the large-mass regime
\begin{equation}\label{Eq16}
l+1\ll M\mu\  .
\end{equation}
In the eikonal large-mass regime (\ref{Eq16}), the radial potential
(\ref{Eq13}), which characterizes the dynamics of the charged
massive scalar field in the charged RN black-hole spacetime, can be
approximated by
\begin{equation}\label{Eq17}
V(r)=\Big(\omega-{{qQ}\over{r}}\Big)^2-\Big(1-{{2M}\over{r}}+{{Q^2}\over{r^2}}\Big)\mu^2\cdot\{1+O[(M\mu)^{-2}]\}.
\end{equation}
This radial potential has the form of an effective potential barrier
whose maximum $r_0$ is located at
\begin{equation}\label{Eq18}
r_0={{Q^2(q^2-\mu^2)}\over{qQ\omega-M\mu^2}}\  .
\end{equation}

As we shall now show, the fundamental complex resonances associated
with the effective scattering potential (\ref{Eq13}) can be
determined analytically in the large-mass regime (\ref{Eq16}) using
standard WKB methods \cite{WKB1,WKB2,WKB3,Will}. In particular, as
shown in \cite{WKB1,WKB2}, the WKB resonance condition which
characterizes the complex scattering resonances (the quasinormal
frequencies) of the Schr\"odinger-like radial equation (\ref{Eq12})
in the eikonal large-frequency regime is given by
\begin{equation}\label{Eq19}
{{iV_0}\over{\sqrt{2V^{(2)}_0}}}=n+{1\over 2}\ ,
\end{equation}
where the various derivatives $V^{(k)}_0\equiv d^{k}V/dy^{k}$ (with
$k\geq0$) that appear in the WKB resonance equation (\ref{Eq19}) are
evaluated at the maximum point $y=y_0(r_0)$ [see Eq. (\ref{Eq18})]
which characterizes the effective scattering potential $V(y)$.

Substituting Eqs. (\ref{Eq17}) and (\ref{Eq18}) into the WKB
resonance equation (\ref{Eq19}) and using the differential relation
(\ref{Eq10}), one finds the characteristic resonance condition
\begin{equation}\label{Eq20}
{{\Big(\omega-{{qQ}\over{r_0}}\Big)^2-\Big(1-{{2M}\over{r_0}}+{{Q^2}\over{r^2_0}}\Big)\mu^2}\over
{2\Big(1-{{2M}\over{r_0}}+{{Q^2}\over{r^2_0}}\Big)
\sqrt{{{qQ\omega-M\mu^2}\over{r^3_0}}}}}= -i\Big(n+{1\over2}\Big)\
\end{equation}
for the quasinormal resonant frequencies of the composed
Reissner-Nordstr\"om-black-hole-charged-massive-scalar-field system.
As we shall now show, the rather complicated resonance equation
(\ref{Eq20}) can be solved {\it analytically} in the regime
\cite{Notewri}
\begin{equation}\label{Eq21}
\omega_{\text{R}}\gg \omega_{\text{I}}\  .
\end{equation}
This strong inequality, which characterizes the fundamental
quasinormal frequencies of the composed black-hole-field system in
the eikonal regime (\ref{Eq2}) [see Eqs. (\ref{Eq25}) and
(\ref{Eq29}) below], enables one to {\it decouple} the real and
imaginary parts of the WKB resonance equation (\ref{Eq20}). In
particular, one finds
\begin{equation}\label{Eq22}
\Big(\omega_{\text{R}}-{{qQ}\over{r_0}}\Big)^2-\Big(1-{{2M}\over{r_0}}+{{Q^2}\over{r^2_0}}\Big)\mu^2=0\
\end{equation}
for the real part of the resonance equation (\ref{Eq20}), and
\begin{equation}\label{Eq23}
\Big(\omega_{\text{R}}-{{qQ}\over{r_0}}\Big)\omega_{\text{I}}=\Big(1-{{2M}\over{r_0}}+{{Q^2}\over{r^2_0}}\Big)
\sqrt{{{qQ\omega_{\text{R}}-M\mu^2}\over{r^3_0}}}\cdot\Big(n+{1\over2}\Big)\
\end{equation}
for the imaginary part of the resonance equation (\ref{Eq20}).

Substituting Eq. (\ref{Eq18}) into Eq. (\ref{Eq22}), one finds
\begin{equation}\label{Eq24}
{{r_0}\over{M}}={{1-{\bar\mu}^2+\sqrt{(1-{\bar\mu}^2)(1-{\bar
Q}^2)}}\over{1-({\bar\mu}/{\bar Q})^2}}
\end{equation}
and
\begin{equation}\label{Eq25}
\omega_{\text{R}}=q{\bar Q}\cdot {{1+\Big({{\bar\mu}\over{\bar
Q}}\Big)^2\sqrt{{{1-{\bar
Q}^2}\over{1-{\bar\mu}^2}}}}\over{1+\sqrt{{{1-{\bar
Q}^2}\over{1-{\bar\mu}^2}}}}}\  ,
\end{equation}
where
\begin{equation}\label{Eq26}
{\bar\mu}\equiv {{\mu}\over{q}}\ \ \ \ ; \ \ \ \ {\bar Q}\equiv
{{Q}\over{M}}\  .
\end{equation}
Note that the relations (\ref{Eq24}) and (\ref{Eq25}) are valid in
the regime ${\bar\mu}<{\bar Q}\leq1$ \cite{Notemql}, which
corresponds to
\begin{equation}\label{Eq27}
{{\mu}\over{q}}<{{Q}\over{M}}\leq 1\  .
\end{equation}

Interestingly, one finds from (\ref{Eq25}) that, for a given value
of the dimensionless black-hole electric charge ${\bar Q}$, the real
part of the resonant oscillation frequency, $\omega_{\text{R}}$, is
a monotonically {\it increasing} function of the dimensionless ratio
${\bar\mu}$. In particular, one finds the limiting behaviors [see
Eq. (\ref{Eq25})]
\begin{equation}\label{Eq28}
\{\omega_{\text{R}}\to {(qQ/r_+)}^+\ \ \ \text{for}\ \ \
{\bar\mu}/{\bar Q}\to0\}\ \ \ \ \ \text{and}\ \ \ \ \
\{\omega_{\text{R}}\to {(qQ/M)}^-\ \ \ \text{for}\ \ \
{\bar\mu}/{\bar Q}\to1^-\}\  .
\end{equation}

Substituting Eqs. (\ref{Eq24}) and (\ref{Eq25}) into Eq.
(\ref{Eq23}), one finds
\begin{equation}\label{Eq29}
M\omega_{\text{I}}=\sqrt{1-{\bar Q}^2}\Big[{{1-({\bar\mu}/{\bar
Q})^2}\over{1-{\bar\mu}^2+\sqrt{(1-{\bar\mu}^2)(1-{\bar
Q}^2)}}}\Big]^2\cdot\Big(n+{{1}\over{2}}\Big)
\end{equation}
for the imaginary parts of the quasinormal resonances which
characterize the composed RN-black-hole-charged-massive-scalar-field
system in the regime (\ref{Eq2}). Interestingly, one finds from
(\ref{Eq29}) that, for a given value of the dimensionless black-hole
electric charge ${\bar Q}$, the imaginary part of the resonant
oscillation frequency, $\omega_{\text{I}}$, is a monotonically {\it
decreasing} function of the dimensionless ratio ${\bar\mu}$. In
particular, one finds the limiting behaviors [see Eq. (\ref{Eq29})]
\begin{equation}\label{Eq30}
\{M\omega_{\text{I}}\to {{\sqrt{1-{\bar Q}^2}}\over{(1+\sqrt{1-{\bar
Q}^2})^2}}\cdot (n+1/2)\ \ \ \text{for}\ \ \ {\bar\mu}/{\bar
Q}\to0\}\ \ \ \ \ \text{and}\ \ \ \ \ \{M\omega_{\text{I}}\to 0^+ \
\ \ \text{for}\ \ \ {\bar\mu}/{\bar Q}\to1^-\}\  .
\end{equation}

Note that the expression (\ref{Eq29}) for the imaginary parts of the
RN-black-hole-charged-massive-scalar-field resonances can be written
in the compact form [see Eqs. (\ref{Eq5}) and (\ref{Eq24})]
\begin{equation}\label{Eq31}
\omega_{\text{I}}=2\pi
T_{\text{BH}}\Big({{r_+}\over{r_0}}\Big)^2\cdot\Big(n+{{1}\over{2}}\Big)\
,
\end{equation}
where
\begin{equation}\label{Eq32}
T_{\text{BH}}={{r_+-r_-}\over{4\pi r^2_+}}\
\end{equation}
is the Bekenstein-Hawking temperature of the RN black hole
\cite{Notebnd,Hodtpi,Noterib}.

\section{The regime of validity of the WKB approximation}

It is important to emphasize that our WKB results (\ref{Eq25}) and
(\ref{Eq29}) for the real and imaginary parts of the quasinormal
resonant frequencies which characterize the relaxation dynamics of
the charged massive scalar fields in the charged RN black-hole
spacetime were derived under the assumption that higher-order
correction terms that appear in the large-frequency WKB
approximation can be neglected. In particular, as explicitly shown
in \cite{WKB1,WKB2,WKB3,Will}, an extension of the WKB approximation
to include higher-order derivatives of the effective scattering
potential yields the correction term
\begin{equation}\label{Eq33}
\Lambda(n)={{1}\over{\sqrt{2V^{(2)}_0}}}\Big[{{1+(2n+1)^2}\over{32}}\cdot{{V^{(4)}_0}\over{V^{(2)}_0}}-
{{28+60(2n+1)^2}\over{1152}}\cdot\Big({{V^{(3)}_0}\over{V^{(2)}_0}}\Big)^2\Big]\
\end{equation}
on the r.h.s of the resonance equation (\ref{Eq19}). Thus, the WKB
resonance condition (\ref{Eq19}) is valid provided
\begin{equation}\label{Eq34}
{{\Lambda(n)}\over{n+{1\over 2}}}\ll1\  .
\end{equation}

Substituting Eqs. (\ref{Eq10}), (\ref{Eq17}), (\ref{Eq24}), and
(\ref{Eq25}) into Eq. (\ref{Eq34}), one finds
\begin{equation}\label{Eq35}
\Lambda(n)={{1}\over{qQ\sqrt{1-{\bar\mu}^2}}}\times
\Big[{1\over8}\Big(1-{{6M}\over{r_0}}+{{6Q^2}\over{r^2_0}}\Big)
-{3\over8}\Big(1-{{10M}\over{3r_0}}+{{10Q^2}\over{3r^2_0}}+{{4(1-{\bar\mu}^2)Q^2}
\over{3{\bar\mu}^2r^2_0}}\Big)\cdot(2n+1)^2\Big]\  ,
\end{equation}
which implies that the WKB resonance condition that we have used,
Eq. (\ref{Eq19}), is valid in the large coupling (eikonal) regime
[see Eq. (\ref{Eq34})] \cite{NoteSchw,Schw}
\begin{equation}\label{Eq36}
qQ\sqrt{1-{\bar\mu}^2}\gg n+{1\over2}\  .
\end{equation}

\section{Summary and Discussion}

We have determined the characteristic timescales associated with the
relaxation dynamics of the composed
Reissner-Nordstr\"om-black-hole-charged-massive-scalar-field system.
To that end, the quasinormal resonance spectrum
$\{\omega_n(\mu,q,M,Q)\}_{n=0}^{n=\infty}$ which characterizes the
dynamics of a linearized charged scalar field of mass $\mu$ and
charge coupling constant $q$ in the charged RN black-hole spacetime
of mass $M$ and electric charge $Q$ was studied {\it analytically}
in the eikonal regime $1\ll M\mu<qQ$ [see Eq. (\ref{Eq2})]. In
particular, we have derived the analytical expression [see Eqs.
(\ref{Eq25}) and (\ref{Eq29})]
\begin{equation}\label{Eq37}
M\omega(M,Q,\mu,q;n)=qQ\cdot {{1+\Big({{\bar\mu}\over{\bar
Q}}\Big)^2\sqrt{{{1-{\bar
Q}^2}\over{1-{\bar\mu}^2}}}}\over{1+\sqrt{{{1-{\bar
Q}^2}\over{1-{\bar\mu}^2}}}}}-i \sqrt{1-{\bar
Q}^2}\Big[{{1-({\bar\mu}/{\bar
Q})^2}\over{1-{\bar\mu}^2+\sqrt{(1-{\bar\mu}^2)(1-{\bar
Q}^2)}}}\Big]^2\cdot\Big(n+{{1}\over{2}}\Big)\
\end{equation}
for the quasinormal resonant frequencies which characterize the
composed RN-black-hole-charged-massive-scalar-field system in the
eikonal regime (\ref{Eq2}) \cite{Noteap,Notepv}.

The characteristic timescale $\tau_{\text{relax}}\equiv
1/\Im\omega_0$ associated with the linearized relaxation dynamics of
the composed RN-black-hole-charged-massive-scalar-field system is
determined by its fundamental (least damped) quasinormal resonant
frequency. In particular, from (\ref{Eq37}) one finds the expression
\begin{equation}\label{Eq38}
{{\tau_{\text{relax}}}\over{M}}={{2}\over{\sqrt{1-{\bar
Q}^2}}}\Big[{{1-{\bar\mu}^2+\sqrt{(1-{\bar\mu}^2)(1-{\bar
Q}^2)}}\over{1-({\bar\mu}/{\bar Q})^2}}\Big]^2
\end{equation}
for the characteristic relaxation time of the composed
black-hole-field system. Interestingly, one finds from (\ref{Eq38})
that the composed RN-black-hole-charged-massive-scalar-field system
is characterized by extremely long relaxation times in the limiting
case ${\bar\mu}/{\bar Q}\to 1^-$:
\begin{equation}\label{Eq39}
\tau_{\text{relax}}\to\infty\ \ \ \ \text{for}\ \ \ \
{{M\mu}\over{qQ}}\to 1^-\  .
\end{equation}
Thus, although Reissner-Nordstr\"om black holes cannot support
static matter fields outside their horizons
\cite{Bek1,Notenhc,Hodth,Sham,Notesham}, we conclude that these
black-hole spacetimes may host extremely {\it long-lived}
(exponentially decaying with long relaxation times,
$\tau_{\text{relax}}\gg M$) charged massive scalar fields in their
external regions.

\bigskip
\noindent
{\bf ACKNOWLEDGMENTS}
\bigskip

This research is supported by the Carmel Science Foundation. I would
like to thank Yael Oren, Arbel M. Ongo, Ayelet B. Lata, and Alona B.
Tea for helpful discussions.



\begin{thebibliography}{99}
\bibitem{Notetim} That is, the timescale it takes for the black hole to swallow
or to scatter away to infinity the external matter fields.
\bibitem{Notetails} It is worth noting that, in asymptotically flat black-hole spacetimes,
these exponentially damped quasinormal oscillations, which
characterize the composed black-hole-field system, are followed by
decaying power-law wave tails \cite{Tails} which reflect the
backscattering of the fields by the effective curvature potential of
the black-hole spacetime at large distances from the central black
hole.
\bibitem{Notenhc} It is important to stress the fact that existing
no-hair theorems (see \cite{Bek1} and references therein) rule out
the existence of asymptotically flat static charged
black-hole-scalar-field hairy configurations. In addition, it was
proved in \cite{Hodth} that stationary (or exponentially growing in
time) charged scalar fields (linearized bound-state charged scalar
configurations) cannot be supported in the external spacetime
regions of asymptotically flat charged Reissner-Nordstr\"om black
holes. Thus, charged scalar fields which propagate outside the outer
horizon of a spherically symmetric asymptotically flat
Reissner-Nordstr\"om black hole are expected to be swallowed by the
black hole or to be scattered away to infinity.
\bibitem{Noteunit} We shall use natural units in which $G=c=\hbar=k_{\text{B}}=1$.
\bibitem{Noteqm} Note that $\mu$ and $q$ stand for $\mu/\hbar$ and $q/\hbar$, respectively.
Hence, these field parameters have the dimensions of
$($length$)^{-1}$.
\bibitem{Notegen} We shall assume $Q>0$ and $q>0$ without loss of
generality.
\bibitem{Noteom} The integer parameters $m$ and $l\geq |m|$ are respectively the
azimuthal harmonic index and the spherical harmonic index which
characterize the scalar field mode. We shall henceforth omit these
harmonic indices for brevity.
\bibitem{Notesy} It is important to note that the functions $S_{lm}(\theta)$
are the familiar associated Legendre polynomials
$Y_l^m(\cos\theta)$.
\bibitem{Notewri} Note that we use the standard convention
$\omega=\omega_{\text{R}}-i\omega_{\text{I}}$ for the characteristic
quasinormal frequencies of the composed black-hole-field system.
\bibitem{Notebnd} It is interesting to note that the relation
(\ref{Eq31}) with $r_0>r_+$ yields the characteristic inequality
$\Im\omega_0<\pi T_{\text{BH}}$ for the fundamental (least damped)
quasinormal resonance of the composed
RN-black-hole-charged-massive-scalar-field system. This
characteristic inequality implies that the relaxation time
$\tau_{\text{relax}}=1/\Im\omega_0$ of the composed black-hole-field
system is bounded from below by the relation
$\tau_{\text{relax}}\times T_{\text{BH}}>1/\pi$, in agreement with
the conjectured time-times-temperature (TTT) inequality
\cite{Hodtpi}.
\bibitem{Noterib} Note that $\omega_{\text{R}}>qQ/r_+$ whereas $\omega_{\text{I}}<2\pi
T_{\text{BH}}\cdot(n+1/2)<(n+1/2)/4M$ [see Eqs. (\ref{Eq28}) and
(\ref{Eq30})]. Thus, our assumption
$\omega_{\text{R}}\gg\omega_{\text{I}}$ [see (\ref{Eq21})] is valid
in the large coupling (eikonal) regime $qQ\gg n+1$ [see Eq.
(\ref{Eq2})].
\bibitem{NoteSchw} It is worth emphasizing the fact that vacuum polarization effects (the quantum
discharge phenomenon of charged black holes caused by the
Schwinger-type pair production mechanism \cite{Schw}) restrict the
physical parameters (masses and electric charges) of the composed
RN-black-hole-charged-massive-scalar-field system to the quantum
regime $qQ\ll\mu^2r^2_+/\hbar$ \cite{Hodch2}. Thus, one finds
$\sqrt{1-{\bar\mu}^2}qQ/\mu^2 r^2_0<qQ/\mu^2 r^2_+\ll1$ for the last
term in Eq. (\ref{Eq35}).
\bibitem{Noteap} Note that, in the extremal $(1-{\bar Q}^2)/(1-{\bar\mu}^2)\to 0\ (T_{\text{BH}}\to0)$ limit,
one finds from Eq. (\ref{Eq37}) the leading-order behavior
$\omega_n=qQ/r_+-i2\pi T_{\text{BH}}\cdot(n+1/2)$ for the
quasinormal resonant frequencies which characterize the composed
near-extremal RN-black-hole-charged-massive-scalar-field system.
This expression agrees with the results presented in \cite{Hodch1}.
\bibitem{Notepv} Note that, in the small mass ${\bar\mu}/{\bar Q}\ll1$ limit,
one finds from Eq. (\ref{Eq37}) the leading-order behavior
$\omega_n=qQ/r_+-i2\pi T_{\text{BH}}\cdot(n+1/2)$ for the
quasinormal resonant frequencies which characterize the composed
RN-black-hole-charged-scalar-field system. This expression agrees
with the results presented in \cite{Hodch2,Konf,Ric}.
\end{thebibliography}
\end{document}